\documentclass[journal]{IEEEtran}

\usepackage[T1]{fontenc}
\usepackage{amsmath,amssymb,amsfonts,bm}
\usepackage{graphicx}
\usepackage{xcolor}
\usepackage{cite}
\usepackage{url}
\usepackage{float}
\usepackage[section]{placeins}   
\usepackage[caption=false,font=footnotesize]{subfig} 
\usepackage[hidelinks]{hyperref} 

\newcommand{\ket}[1]{\left|#1\right\rangle}
\newcommand{\bra}[1]{\left\langle #1\right|}
\newcommand{\ev}[1]{\left\langle #1\right\rangle}
\newcommand{\Jrat}{J_2/J_1}

\newcommand{\figcol}[3][4.2cm]{%
  \IfFileExists{#2}{\includegraphics[width=\columnwidth]{#2}}%
  {\fbox{\parbox[c][#1][c]{0.9\columnwidth}{\centering
  \textbf{Figure placeholder}\\[3pt]{\small\texttt{\detokenize{#2}}}\\[3pt]{\footnotesize #3}}}}}

\newcommand{\panel}[3][5cm]{%
  \begin{minipage}[b]{0.48\textwidth}
    \centering
    \IfFileExists{#2}{\includegraphics[width=\linewidth]{#2}}%
    {\fbox{\parbox[c][#1][c]{0.9\linewidth}{\centering
      \textbf{Figure placeholder}\\[3pt]{\small\texttt{\detokenize{#2}}}}}}
    \par\vspace{3pt}{\footnotesize #3\par}
  \end{minipage}}

\begin{document}

\title{\Large Depth-Dependent Collapse and Warm-Start Limits of a Symmetry-Preserving VQE for the J$_1$--J$_2$ Heisenberg Ring}

\author{Soumyadip~Sahoo,~Sahinur~Reja%
\thanks{S.~Sahoo and S.~Reja are with the Department of Physics, Jadavpur University, Kolkata 700032, India (e-mail: soumyadipsahoo867@gmail.com).}%
\thanks{November 2026.}}

\markboth{}{Sahoo and Reja: Collapse and Warm-Start Limits of a Symmetry-Preserving VQE}

\maketitle


\begin{abstract}
Increasing circuit depth is expected to improve a variational quantum eigensolver (VQE). We show that this expectation can fail, and that curing the failure does not guarantee the exact ground state. We study a symmetry-preserving ansatz of SingleExcitationMinus gates, started from the N\'eel state, on the frustrated spin-$1/2$ $J_1$--$J_2$ Heisenberg ring with $N=10$ and $12$ sites. With random initialization the fidelity, energy error, half-chain entropy and total spin change abruptly at intermediate depth ($L\approx 9$ for $N=10$, $L=10$--$12$ for $N=12$), giving improvement, degradation and recovery, a double-descent-like shape. The circuit family is nested and the gradient stays finite, so this is an optimization failure, not a loss of expressibility or a barren plateau. A chained warm start, which extends an optimized shallow circuit with identity layers, removes the collapse and improves energy error and fidelity almost monotonically for most frustration values. However, near $J_2/J_1=0.60$ ($N=10$) and over $J_2/J_1=0.60$--$1.00$ ($N=12$) the warm start stalls at a fidelity of $\approx 0.50$--$0.58$, although the energy error is small and the total spin stays in the $S=0$ sector. Removing the non-monotonic depth dependence is therefore not the same as reaching the ground state, and low energy and a singlet-like state are not sufficient: the fidelity, entropy and total spin must be satisfied together. The conclusions are limited to the system sizes and frustration values studied.
\end{abstract}


\section{Introduction}
\IEEEPARstart{N}{oisy} intermediate-scale quantum (NISQ) devices~\cite{Preskill2018} have made variational quantum algorithms~\cite{Cerezo2021,Tilly2022} a promising route to the Hamiltonian simulation of quantum many-body systems. The variational quantum eigensolver (VQE)~\cite{Peruzzo2014} converts ground-state estimation into an optimization problem over a parameterized quantum circuit, which is solved with a classical optimizer. Increasing the circuit depth enlarges the accessible state manifold and is therefore expected to improve the result. The practical behavior is, however, governed by the interplay of state-space reachability, the geometry of the objective landscape and the classical optimizer used to explore it. These ingredients must be kept distinct: barren plateaus concern the concentration of gradient information~\cite{McClean2018,Cerezo2021b}, whereas poor optimization can also arise from competing minima, parameter redundancy or unfavorable initialization~\cite{Anschuetz2022,Wierichs2020}.

Several strategies address these difficulties. Identity-block initialization~\cite{Grant2019} and layerwise training~\cite{Skolik2021} start from circuits close to the identity, and warm-start and continuation strategies have been investigated through the transferability of smooth variational solutions, generative flow-based initialization and adiabatic deformation~\cite{Mele2022,Zou2026,Puig2026}. Lie-algebraic and overparameterization theories provide a useful language for the relation between parameter count, reachable manifolds and trainability~\cite{Ragone2024,Fontana2024,Larocca2023}, and recent work has analyzed the geometry of VQE optimization, including symmetry-protected basin localization, optimization-landscape geometry in frustrated spin models and geometric analyses of VQE~\cite{Wang2026,Novak2026,Qin2026}.

A second motivation is the recent observation of double-descent-like behavior in quantum learning. In classical machine learning the term describes a non-monotonic dependence of the test error on model complexity, most famously through an interpolation regime~\cite{Belkin2019,Nakkiran2019}. In quantum settings, Kempkes \emph{et al.} demonstrated double descent in quantum kernel methods~\cite{Kempkes2026}, and Moss \emph{et al.} reported related behavior in neural quantum states~\cite{Moss2026}. A VQE is a different problem: it has a fixed physical target rather than a train/test split. The question addressed here is therefore narrower and more physical: \emph{can increasing the circuit depth temporarily worsen the quality of a VQE ground-state approximation before it recovers?}

Frustrated quantum magnets are a demanding test for ground-state algorithms~\cite{Balents2010}. Competing interactions suppress simple magnetic order, produce small spectral gaps and quasi-degenerate low-energy manifolds, and invalidate the sign structure that makes some classical methods work for unfrustrated lattices. The spin-$1/2$ $J_1$--$J_2$ Heisenberg model, with antiferromagnetic nearest-neighbor (NN) coupling $J_1$ and next-nearest-neighbor (NNN) coupling $J_2$, is a canonical example. In one dimension it hosts a quantum phase transition to a spontaneously dimerized phase at $J_2/J_1\approx 0.24$~\cite{Haldane1982,Eggert1996,Okamoto1992}, the exactly solvable Majumdar--Ghosh point at $J_2/J_1=0.5$~\cite{Majumdar1969}, and incommensurate spiral correlations at larger $J_2/J_1$~\cite{White1996}. The model therefore combines a nontrivial ground-state structure with an exactly tractable small-system benchmark, which makes it a useful setting for the question above.

Prior VQE studies have investigated ansatz construction for the square-lattice $J_1$--$J_2$ model~\cite{Feulner2022}, the VQE behavior across a phase-transition point and related Heisenberg problems on the kagome lattice~\cite{Guo2023,Kattemolle2022,Bosse2022}, and near-term simulation of the square-lattice spin-$1/2$ model~\cite{Sheils2024}. Brooks \emph{et al.} recently explored frustrated Heisenberg systems with sample-based Krylov quantum diagonalization, a complementary route to ground-state estimation~\cite{Brooks2026}. Symmetry-adapted VQE~\cite{Seki2020} and the Hamiltonian variational ansatz~\cite{Wiersema2020} treat the symmetries of the Heisenberg model explicitly. We instead study a periodic one-dimensional ring and focus on the optimization behavior of a symmetry-preserving ansatz built from number-conserving two-qubit gates~\cite{Arrazola2022}.

The purpose of the present work is not to claim the first use of the phrase ``double descent'' in a quantum algorithm. Instead, we characterize a depth-dependent performance collapse in a symmetry-preserving VQE for a frustrated $J_1$--$J_2$ ring, distinguish it from a barren plateau, and test whether a better solution can be recovered by a simple within-system warm-start continuation. The central empirical object is the optimizer's outcome as the depth is changed. The energy error, the fidelity, the half-chain entropy, the total spin and the gradient are used to determine what does \emph{not} explain the collapse and which changes of the state accompany it. To our knowledge, the behavior of warm-start continuation across the frustration range of a symmetry-preserving $J_1$--$J_2$ ring has not been mapped yet.

Our main results are as follows.
\begin{enumerate}
  \item For random initialization at $N=10$, the fidelity, the energy error, the half-chain entropy and the total spin change abruptly at a depth of about nine and recover gradually at larger depth.
  \item The gradient stays finite at the abrupt change, which excludes a barren plateau.
  \item A chained warm start removes the abrupt change and gives a monotonic improvement of the energy error and of the fidelity for most frustration values.
  \item Near $J_2/J_1=0.6$ the warm start fails to reach the ground state for $N=10$ although it stays in the $S=0$ sector, which shows that several conditions must hold simultaneously for the exact ground state.
  \item For $N=12$ the abrupt change moves to $10$--$12$ layers and is stronger, with no recovery to high fidelity up to $16$ layers, and the warm-start failure widens from a single frustration value to the range $J_2/J_1=0.6$--$1.0$.
\end{enumerate}

The paper is organized as follows. Section~\ref{sec:model} defines the model, the ansatz, the observables and the numerical implementation. Section~\ref{sec:results} presents the results and their discussion for $N=10$ and $N=12$. Section~\ref{sec:limits} gives the limitations and an outlook, and Section~\ref{sec:concl} concludes.

\section{Model and Variational Problem}\label{sec:model}

\subsection{\texorpdfstring{J$_1$--J$_2$}{J1--J2} Heisenberg ring}
\begin{figure}[H]
  \centering
  \figcol[4.5cm]{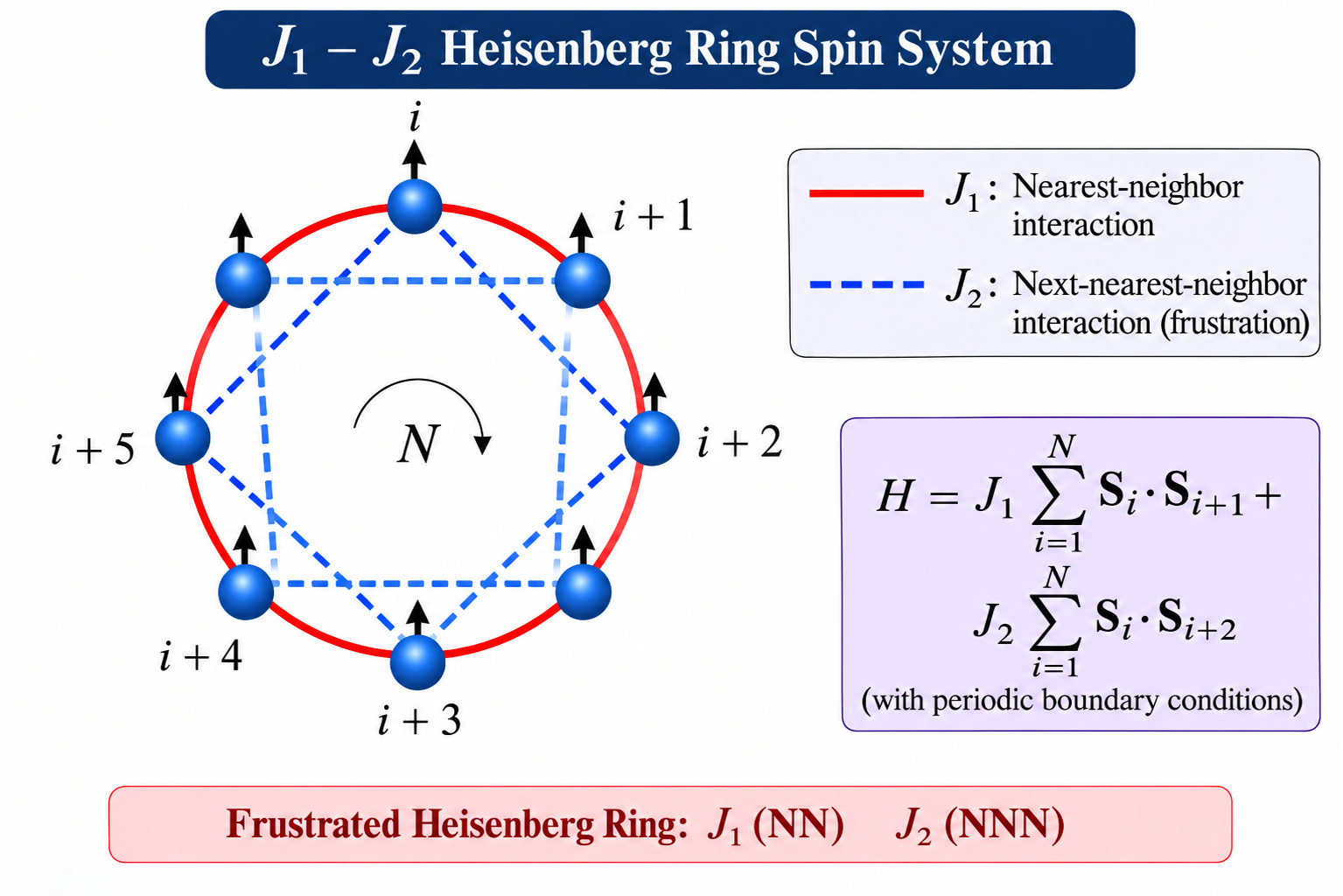}{Ring of $N$ spins with NN coupling $J_1$ and NNN coupling $J_2$ and periodic boundary conditions.}
  \caption{Frustrated Heisenberg ring: NN coupling $J_1$ and NNN coupling $J_2$ on a periodic chain of $N$ spin-$1/2$ sites.}
  \label{fig:ring}
\end{figure}

We consider an even periodic chain of $N$ spin-$1/2$ degrees of freedom (Fig.~\ref{fig:ring}),
\begin{equation}
  H = J_1\sum_{i=0}^{N-1}\boldsymbol{\sigma}_i\!\cdot\!\boldsymbol{\sigma}_{i+1}
    + J_2\sum_{i=0}^{N-1}\boldsymbol{\sigma}_i\!\cdot\!\boldsymbol{\sigma}_{i+2},
  \label{eq:H}
\end{equation}
with $\boldsymbol{\sigma}_i\!\cdot\!\boldsymbol{\sigma}_j=X_iX_j+Y_iY_j+Z_iZ_j$ built from Pauli matrices (four times the corresponding expression in spin-$1/2$ operators $\bm S_i=\bm\sigma_i/2$), indices taken modulo $N$, and $J_1=1$ setting the energy scale. The frustration coordinate is $\Jrat$. The calculations use periodic boundary conditions and preserve the total magnetization,
\begin{equation}
  S^z_{\rm tot}=\sum_i S^z_i,\qquad [H,S^z_{\rm tot}]=0.
  \label{eq:Sz}
\end{equation}
The Hamiltonian also commutes with the total spin $\mathbf{S}^2$, so eigenstates are labeled by $(S,S^z)$. The $S^z=0$ sector has dimension $\binom{N}{N/2}$, i.e.\ $252$ for $N=10$ and $924$ for $N=12$. We study $N=10$ and $N=12$ at $\Jrat=0.01,\,0.20,\,0.40,\,0.60,\,0.80,\,1.00$.

\subsection{Symmetry-preserving ansatz}\label{sec:ansatz}
The variational state is
\begin{equation}
  \ket{\psi(\boldsymbol{\theta})}=U(\boldsymbol{\theta})\ket{\psi_{\rm N}},
  \qquad E(\boldsymbol{\theta})=\bra{\psi(\boldsymbol{\theta})}H\ket{\psi(\boldsymbol{\theta})},
  \label{eq:ansatz}
\end{equation}
where $\ket{\psi_{\rm N}}=\ket{\uparrow\downarrow\uparrow\downarrow\cdots}$ is the N\'eel-type product state in the $S^z=0$ sector. Each layer applies the gate to the $N$ NN pairs and then to the $N$ NNN pairs. With one trainable angle per pair, the number of parameters at depth $L$ is
\begin{equation}
  P(L)=2NL.
  \label{eq:P}
\end{equation}
Each pair is acted on by a SingleExcitationMinus gate~\cite{Bergholm2018,Arrazola2022}, which in the basis $\{\ket{00},\ket{01},\ket{10},\ket{11}\}$ of the two qubits is
\begin{equation}
  G(\theta)=
  \begin{pmatrix}
    e^{-i\theta/2} & 0 & 0 & 0\\
    0 & \cos\frac{\theta}{2} & -\sin\frac{\theta}{2} & 0\\
    0 & \sin\frac{\theta}{2} & \cos\frac{\theta}{2} & 0\\
    0 & 0 & 0 & e^{-i\theta/2}
  \end{pmatrix}.
  \label{eq:gate}
\end{equation}
Three properties of $G$ matter for this work.
\begin{itemize}
  \item \emph{U(1) symmetry:} The gate preserves the excitation number and hence $S^z$, so the state stays in the $S^z=0$ sector.
  \item \emph{No SU(2) symmetry:} The gate does not conserve the total spin $\mathbf{S}^2$. The N\'eel reference is not an eigenstate of $\mathbf{S}^2$ and contains all total-spin sectors, and the ansatz can transfer weight between these sectors, including the $S=0$, $S^z=0$ sector in which the exact ground state lies. Its generator is antisymmetric (Dzyaloshinskii--Moriya-like), not the Heisenberg exchange; at $\theta=\pi$ it maps the two-site singlet to the $T_0$ triplet.
  \item \emph{Identity at zero angle:} $G(0)$ is the identity. A new layer with all angles zero leaves the state unchanged, so the circuit family is nested and the optimal energy cannot increase with depth. Any loss of quality with depth is therefore an optimization failure, not a loss of representability.
\end{itemize}
All results use the N\'eel state as the reference state.

\subsection{Warm-start continuation}\label{sec:warm}
If an optimized parameter matrix $\Theta^\star_L$ at depth $L$ is available and the next depth is $L+\Delta L$, the warm start appends $\Delta L$ layers with zero angles (i.e.\ the identity operator) and reoptimizes,
\begin{equation}
  \Theta^{(0)}_{L+\Delta L}=
  \begin{bmatrix}
    \Theta^\star_L\\
    0\\[-0.25em]\vdots\\
    0
  \end{bmatrix}.
  \label{eq:warm}
\end{equation}
The continuation is chained: the circuit at each depth of the grid starts from the optimized solution of the previous grid depth, not directly from the original shallow solution. Because the appended layers are the identity, the initial energy at the new depth equals the optimized energy at the previous depth. The warm start is therefore an intervention on the initialization, not a different ansatz.

\subsection{Exact reference and observables}
Exact diagonalization (ED) of the Hamiltonian~\eqref{eq:H} in the $S^z=0$ sector gives the ground-state energy $E_0$, the ground state $\ket{\psi_0}$, the excitation gap and the exact values of all observables below. The quality of a variational state $\ket{\psi(\boldsymbol{\theta})}$ is measured by two quantities. The energy error is
\begin{equation}
  \Delta E=E(\boldsymbol{\theta})-E_0\ \ge 0,
  \label{eq:dE}
\end{equation}
in units of $J_1$, which is non-negative by the variational principle and zero only for a ground state. The fidelity is the squared overlap with the exact ground state,
\begin{equation}
  F=\bigl|\langle\psi_0|\psi(\boldsymbol{\theta})\rangle\bigr|^2 ,
  \label{eq:F}
\end{equation}
which equals one only if $\ket{\psi(\boldsymbol{\theta})}$ is the ground state~\cite{Nielsen2010}.

The half-chain entanglement entropy~\cite{Amico2008} is obtained by reshaping $\ket{\psi}$ into a bipartite matrix and taking its singular values $s_a$,
\begin{equation}
  S_{\rm vN}=-\mathrm{Tr}\,(\rho_A\ln\rho_A)=-\sum_a s_a^2\ln s_a^2 ,
  \label{eq:S}
\end{equation}
where $\rho_A$ is the reduced density matrix of half the ring. It is computed for the variational state and for $\ket{\psi_0}$, and it serves as a diagnostic of the composition of the state, not as an observable that the optimizer minimizes.

The total spin is evaluated from
\begin{equation}
  \mathbf{S}^2=\sum_{\alpha=x,y,z}\Bigl(\sum_i S^\alpha_i\Bigr)^2
  \label{eq:S2}
\end{equation}
as $\ev{\mathbf{S}^2}$ on the optimized state. For the antiferromagnetic ground state the exact value is zero. A low value indicates proximity to the singlet sector, but a nonzero value in an approximate variational state is not by itself proof of an optimization failure.

Finally, the gradient of the energy with respect to the parameters is used to test for a barren plateau: we report the gradient variance over random parameter samples and the gradient norm at the optimized parameters.

\subsection{Training protocols and numerical implementation}\label{sec:impl}
We simulate the $10$- and $12$-qubit systems with PennyLane~\cite{Bergholm2018} (\texttt{default.qubit}) using noiseless exact statevectors, and minimize $E(\boldsymbol{\theta})$ with the Adam optimizer~\cite{Kingma2015}. Gradients are obtained by automatic differentiation; JAX~\cite{Bradbury2018} is used for statevectors and diagnostics. The initial step size is $0.05$, multiplied by $0.99$ every $100$ iterations, and every run uses a fixed $1500$ iterations with no early stopping. The depth grid is $L=2,4,6,8,9,10,11,12,13,14,16$.

\emph{Random initialization:} Each angle is drawn uniformly from $[0,2\pi)$, with $15$ independent seeds for every frustration value and depth. The random branch shown in all figures is the seed with the lowest final energy (the best seed).

\emph{Chained warm start:} At $L=2$ the warm start begins from the best random run. At each later depth it starts from the optimized parameters of the previous depth, extended by layers with exactly zero angles (Eq.~\eqref{eq:warm}), and is trained for $1500$ iterations with a fresh optimizer.

\emph{Exact reference:} $E_0$ and $\ket{\psi_0}$ come from full diagonalization of $H$; the ground state lies in the $S^z=0$ sector and is non-degenerate for the values shown.

\emph{Gradient diagnostics:} At each depth we compute $\nabla E$ at $30$ random parameter sets and report the component variance, averaged over components, and $\|\nabla E\|$ at the optimized parameters.
\section{Results and Discussion}\label{sec:results}

\subsection{\texorpdfstring{$N=10$}{N=10}: depth-induced collapse with random initialization}
\begin{figure*}[!t]
  \centering
  \subfloat[Behavior of the energy error for the 10-site periodic system at different frustration levels.]{%
    \includegraphics[width=0.95\textwidth]{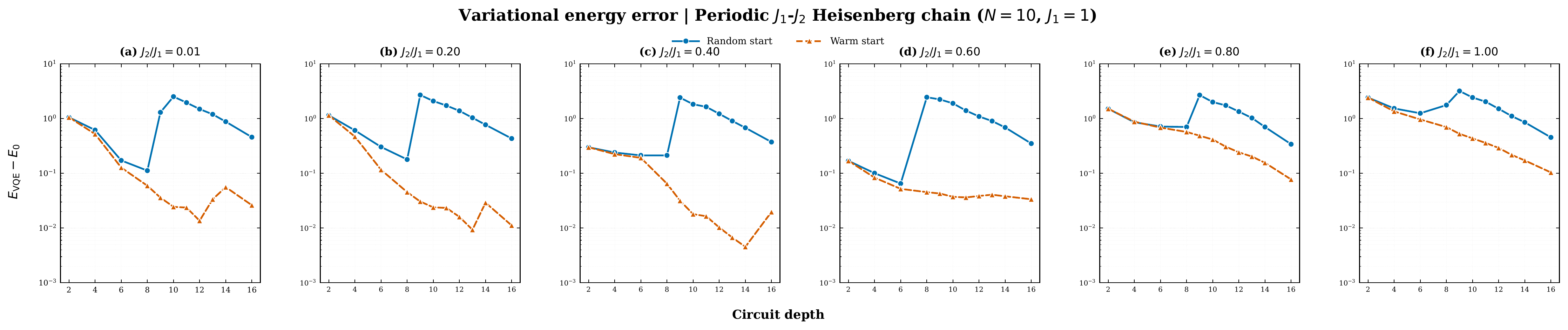}}
  \par\vspace{4mm}
  \subfloat[Behavior of the fidelity for the 10-site periodic system at different frustration levels.]{%
    \includegraphics[width=0.95\textwidth]{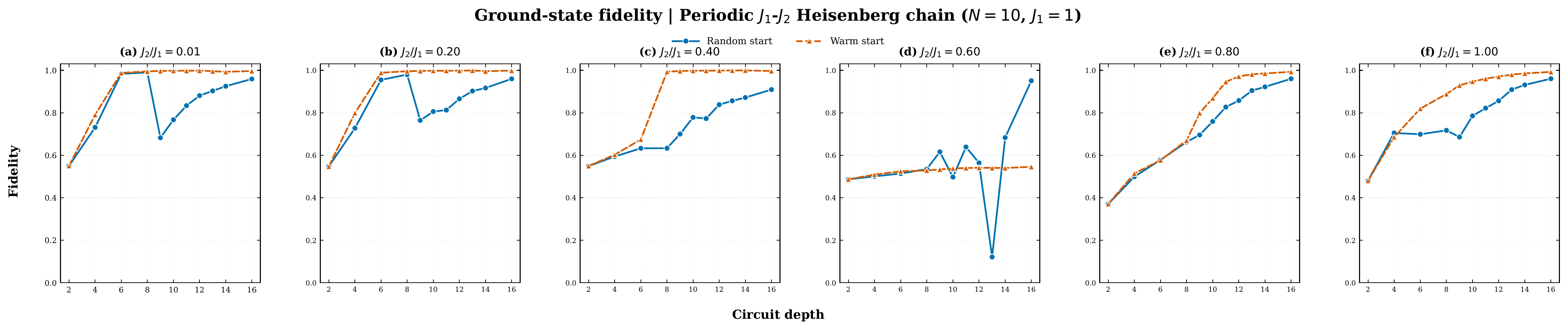}}
  \caption{Energy error and ground-state fidelity versus depth for random initialization and the chained warm start, $N=10$.}
  \label{fig:N10EF}
\end{figure*}

We tested the random-seed behavior of the system with 15 seeds (code and data: \href{https://github.com/soumyadip-sahoo/Heisenberg_J1_J2_VQE}{GitHub}), and the figures show one randomly selected seed. At a depth of about nine, for $\Jrat=0.2$, $0.4$ and $0.8$, the 10-site system shows an abrupt change of the fidelity and of the energy error (Fig.~\ref{fig:N10EF}). The same abrupt change appears in the half-chain entropy and in the total spin (Fig.~\ref{fig:N10SS}).

Up to a depth of $L=8$ the energy error decreases with depth and the energy approaches the exact ground-state energy $E_0$, with high fidelity and a small energy error, for example $\Delta E\approx 0.112$ at $\Jrat=0.01$. Increasing the number of layers increases the number of trainable parameters and thereby the expressibility of the ansatz, which lets the circuit follow the ground state more closely. As the frustration increases, a more expressive ansatz is needed because the ground state becomes more complex.

At the abrupt change the expected trend reverses: instead of a continuous decrease of the energy error and a continuous rise of the fidelity, the energy error increases and, for low frustration value, the fidelity falls to approximately $60\%$. The exception is $\Jrat=0.6$, where the fidelity falls to about $10\%$, the worst performance observed. At $\Jrat=0.8$ the fidelity is largely preserved in spite of the spike of the energy error. At larger depth the system recovers gradually as layers are added; at $L=16$ the fidelity is $0.95$--$0.96$. The peak energy error is $\approx 2.5$. The sequence of improvement, degradation at intermediate depth and recovery resembles double descent~\cite{Belkin2019,Nakkiran2019}, which we use only as a descriptive analogy.

\subsection{\texorpdfstring{$N=10$}{N=10}: entropy and total spin}
\begin{figure*}[!t]
  \centering
  \subfloat[Behavior of the half-chain entropy for the 10-site periodic system at different frustration levels; dashed line: exact value.]{%
    \includegraphics[width=0.95\textwidth]{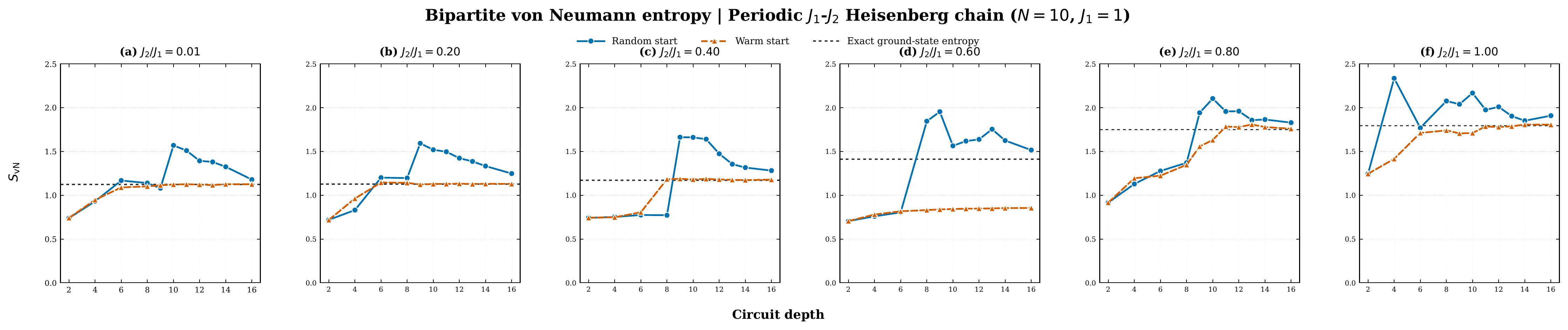}}
  \par\vspace{4mm}
  \subfloat[Behavior of the total spin $\ev{\mathbf{S}^2}$ for the 10-site periodic system at different frustration levels; dashed line: exact value (zero).]{%
    \includegraphics[width=0.95\textwidth]{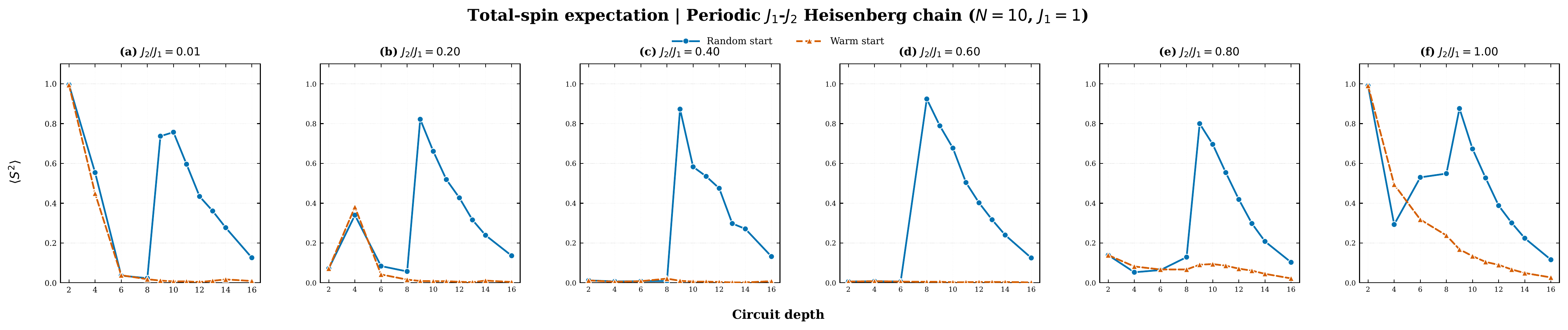}}
  \caption{Half-chain entropy and total spin versus depth for random initialization and the chained warm start, $N=10$. Dashed lines: exact ground-state values.}
  \label{fig:N10SS}
\end{figure*}

The dashed lines in Fig.~\ref{fig:N10SS} give the exact half-chain entropy of the ground state and the exact total spin, which is zero. For small $\Jrat$ the random-initialization branch reaches the ground-state entropy with a small number of layers and moves towards the $\ev{\mathbf{S}^2}=0$ sector. For larger $\Jrat$ the entropy no longer oscillates around the exact value and $\ev{\mathbf{S}^2}$ stays finite, which moves the state away from the singlet ground state; for example, at $\Jrat=0.8$ and depth nine the entropy reaches $1.924$ against the exact value $1.750$, and $\ev{\mathbf{S}^2}$ reaches $0.8$. Since the gates of the ansatz do not conserve $\mathbf{S}^2$, a finite $\ev{\mathbf{S}^2}$ means weight in $S>0$ sectors, which this ansatz can generate. The total spin peaks at $\approx 0.8$--$0.9$. The abrupt change of the energy error and of the fidelity is accompanied by an abrupt change of the entropy and of the total spin, so the four observables describe the same event in different ways.

\subsection{\texorpdfstring{$N=10$}{N=10}: the abrupt change is not a barren plateau}
\begin{figure}[t]
    \centering
    \includegraphics[width=0.95\columnwidth]{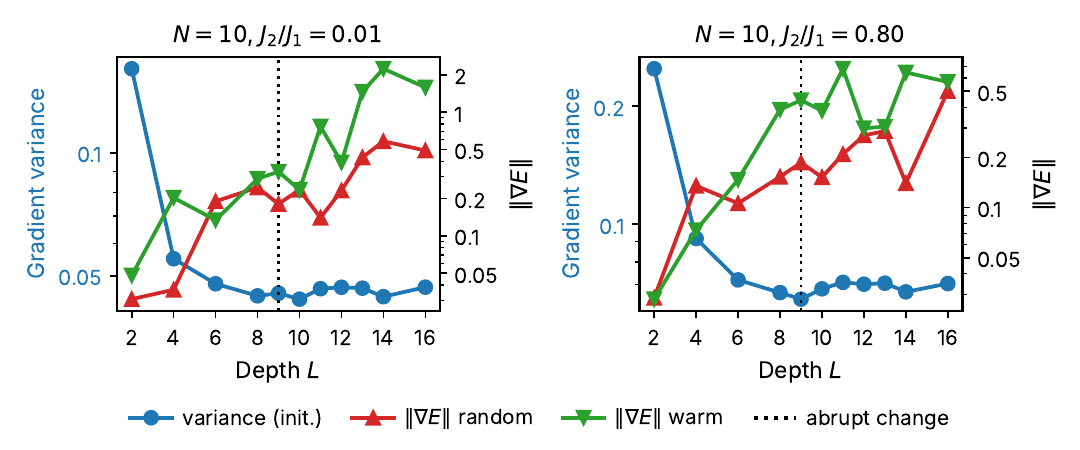}
    \caption{Gradient variance and $\|\nabla E\|$ versus depth ($N=10$). Gradient diagnostics for $N=10$ at $\Jrat=0.01$ (left) and $0.80$ (right). Blue: gradient component variance at random initialization. Red and green: $\|\nabla E\|$ at the optimized random and warm start runs. Dotted line: depth of the abrupt change.}
    \label{fig:N10grad}
\end{figure}

The abrupt change does not mean that the optimization reaches a barren plateau~\cite{McClean2018,Cerezo2021b}. Figure~\ref{fig:N10grad} shows the gradient-component variance at random initialization and $\|\nabla E\|$ at the optimized parameters for $\Jrat=0.01$ and $0.80$. After a drop at small depth, the variance stays constant at $\approx(4.4$--$7.2)\times10^{-2}$ for $L\ge 6$, with no decay with depth and no feature at the abrupt change ($L=9$). At the optimized parameters, $\|\nabla E\|$ stays finite ($0.03$--$2.2$) both before and after the collapse. No run reaches $\|\nabla E\|<10^{-3}$ within $1500$ iterations, so the collapsed states are not shown to be converged minima.

\subsection{\texorpdfstring{$N=10$}{N=10}: the chained warm start}
The chained warm start removes the abrupt change to a large extent. In Figs.~\ref{fig:N10EF} and~\ref{fig:N10SS} the warm-start branch shows a monotonic decrease of the energy error and a monotonic improvement of the fidelity with depth for most frustration values. For $\Jrat=0.40$ the fidelity is already $\approx 0.99$ at $L=8$, and for $\Jrat=0.80$ and $1.00$ it also reaches $\approx 0.99$ at $L=16$. Its entropy oscillates around the exact ground-state entropy and its total spin stays close to zero, i.e.\ in the sector in which the exact ground state lies. The warm start is an intervention that tests accessibility: because the circuit family is nested, the better state reached by the warm start exists at the same depth in the same ansatz with the same expressibility, so the poor random outcome is not forced by representability.

\subsection{\texorpdfstring{$N=10$}{N=10}: failure near \texorpdfstring{$\Jrat=0.6$}{J2/J1=0.6}}
This behavior is not general. Near $\Jrat=0.6$, close to the Majumdar--Ghosh point~\cite{Majumdar1969}, the chained warm start fails to reach the exact ground-state energy $E_0$ (Fig.~\ref{fig:N10EF}). Its fidelity saturates at about $0.54$ and stays there over the whole depth range, and its entropy remains far from the ground-state entropy, while the total spin stays in the $S=0$ sector. The random-initialization branch at this frustration value is erratic, with a deep transient dip, and reaches high fidelity only at the largest depths; at $L=16$ it overtakes the warm start.

This is a clear indication that all conditions must be satisfied simultaneously to obtain the exact ground state. A state in the $S=0$ sector with a small energy error can match several properties of the ground state without being the ground state, so such properties alone say nothing about the low-energy state of the system. Near the Majumdar--Ghosh point the low-energy spectrum becomes nearly degenerate.

\FloatBarrier
\subsection{\texorpdfstring{$N=12$}{N=12}: depth-induced collapse with random initialization}
\begin{figure*}[!t]
  \centering
  \subfloat[Behavior of the energy error for the 12-site periodic system at different frustration levels.]{%
    \includegraphics[width=0.95\textwidth]{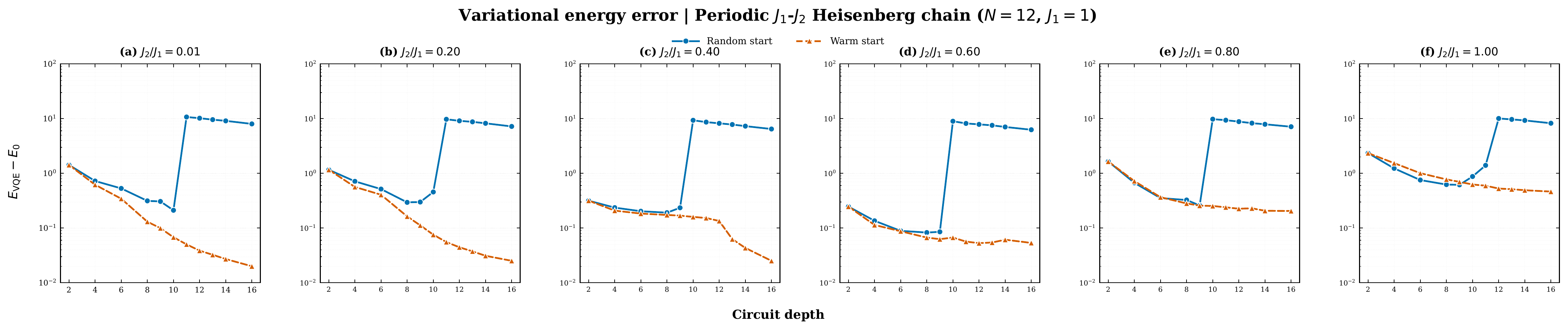}}
  \par\vspace{4mm}
  \subfloat[Behavior of the fidelity for the 12-site periodic system at different frustration levels.]{%
    \includegraphics[width=0.95\textwidth]{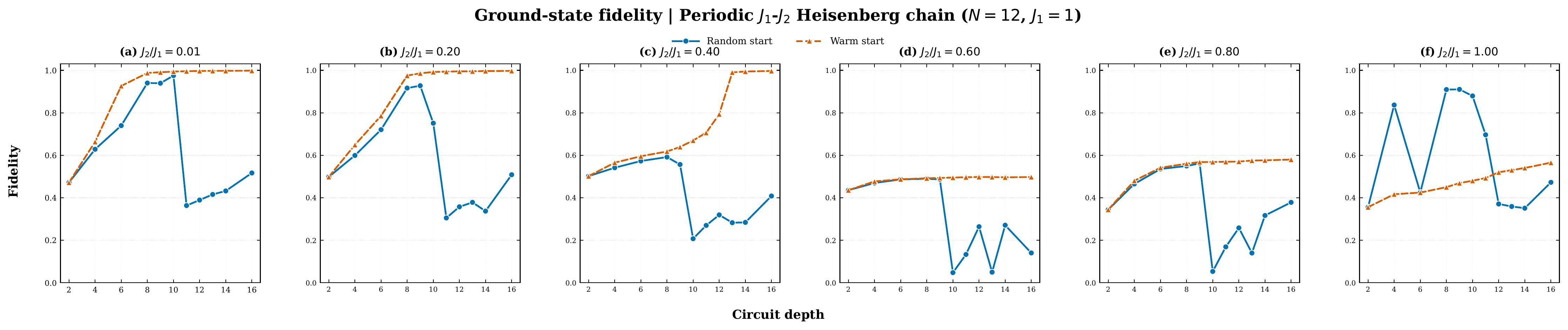}}
  \caption{Energy error and ground-state fidelity versus depth for random initialization and the chained warm start, $N=12$.}
  \label{fig:N12EF}
\end{figure*}

For $N=12$ the $S^z=0$ sector has dimension $924$ and the ansatz has $P(L)=24L$ parameters. As for $N=10$, the random-initialization branch shows an abrupt change of the fidelity and of the energy error (Fig.~\ref{fig:N12EF}), accompanied by an abrupt change of the half-chain entropy and of the total spin (Fig.~\ref{fig:N12SS}).

The abrupt change sits between $L=10$ and $L=11$ for $\Jrat=0.01$ and $0.20$, between $L=9$ and $L=10$ for $\Jrat=0.40$, $0.60$ and $0.80$, and is delayed to $L=12$ for $\Jrat=1.00$, where the energy error already starts to rise from $L=10$. The abrupt change therefore occurs one to three layers later than for $N=10$, where it appeared at $L\approx 9$. In terms of parameters it occurs at $P=240$--$288$ for $N=12$ and at $P=180$ for $N=10$, so it is not tied to a fixed parameter count and also depends on the system size.

Before the abrupt change the energy error decreases with depth and reaches a minimum between $\approx 0.08$ and $\approx 0.6$ just before the jump. The fidelity becomes high only for some frustration values: $\approx 0.97$ at $L=10$ for $\Jrat=0.01$, $\approx 0.93$ at $L=9$ for $\Jrat=0.20$, and $\approx 0.91$ at $L=8$--$9$ for $\Jrat=1.00$. For $\Jrat=0.40$, $0.60$ and $0.80$ the random branch stays at a fidelity of $\approx 0.5$--$0.6$ already before the jump.

At the abrupt change the expected trend reverses. The energy error increases by one to two orders of magnitude, to $\approx 7$--$10$, and the fidelity falls to $\approx 0.3$--$0.4$ for $\Jrat=0.01$, $0.20$ and $1.00$, to $\approx 0.2$ for $\Jrat=0.40$, and to $\approx 0.05$ for $\Jrat=0.60$ and $0.80$, which is the worst performance observed. At larger depth the energy error decreases and the fidelity rises again, but slowly. At $L=16$ the energy error is still $\approx 6$--$8$ and the fidelity is only $0.14$--$0.52$ (for $\Jrat=0.60$ the branch is erratic, with dips to $\approx 0.05$ at $L=13$). In contrast to $N=10$, where the fidelity recovered to $0.91$--$0.96$ at $L=16$, the recovery for $N=12$ has only started at the largest depth studied, and we cannot tell whether it completes at larger depth. The sequence of improvement, degradation at intermediate depth and partial recovery still resembles double descent.

\subsection{\texorpdfstring{$N=12$}{N=12}: entropy and total spin}
\begin{figure*}[!t]
  \centering
  \subfloat[Behavior of the half-chain entropy for the 12-site periodic system at different frustration levels; dashed line: exact value.]{%
    \includegraphics[width=0.95\textwidth]{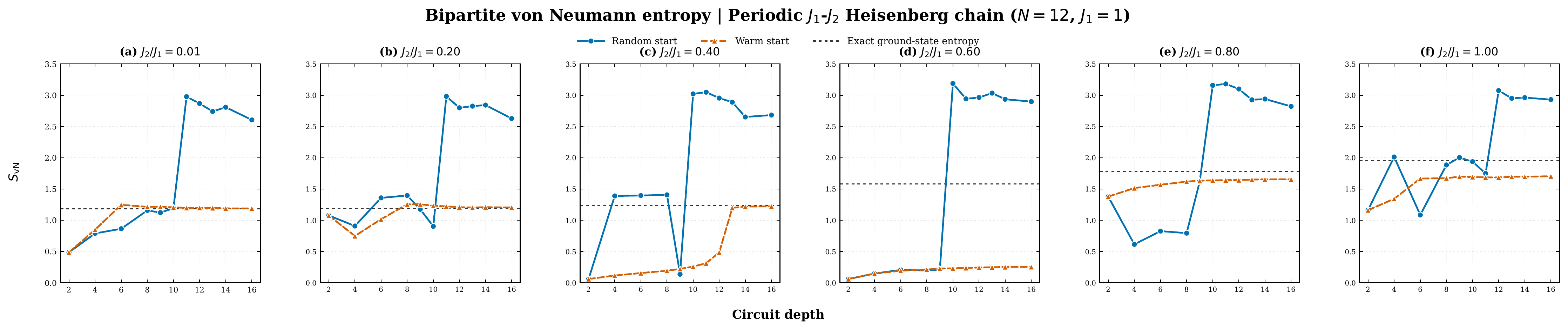}}
  \par\vspace{4mm}
  \subfloat[Behavior of the total spin $\ev{\mathbf{S}^2}$ for the 12-site periodic system at different frustration levels; dashed line: exact value (zero).]{%
    \includegraphics[width=0.95\textwidth]{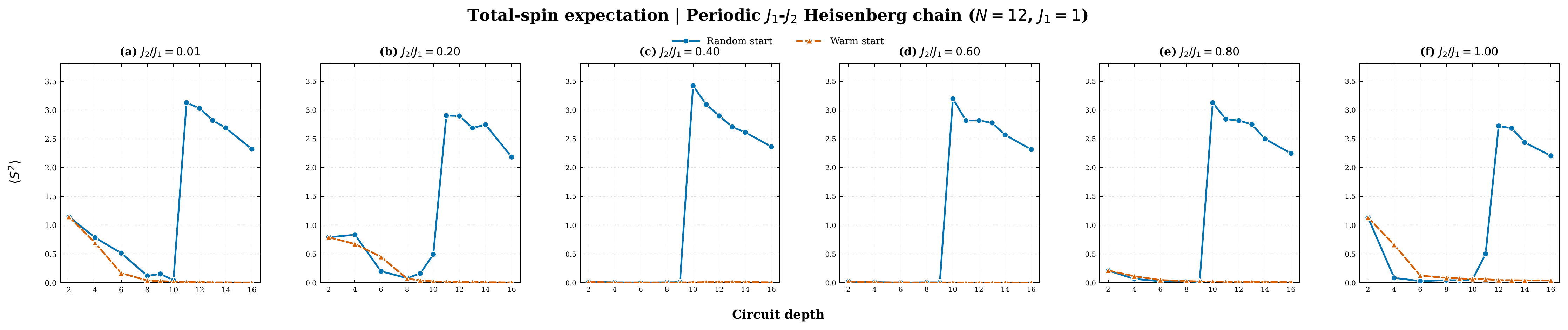}}
  \caption{Half-chain entropy and total spin versus depth for random initialization and the chained warm start, $N=12$. Dashed lines: exact ground-state values.}
  \label{fig:N12SS}
\end{figure*}

The dashed lines in Fig.~\ref{fig:N12SS} give the exact half-chain entropy of the ground state, $S_{\mathrm{vN}}\approx 1.18,\,1.18,\,1.23,\,1.58,\,1.78,\,1.95$ for the six frustration values consecutively, and the exact total spin, which is zero. Before the abrupt change the random branch is close to the singlet sector: $\ev{\mathbf{S}^2}\lesssim 0.2$ for $L\le 8$ at $\Jrat=0.01$ and $0.20$, and $\ev{\mathbf{S}^2}\approx 0$ at $\Jrat=0.40$, $0.60$ and $0.80$. Its entropy is near the exact value for $\Jrat=0.01$ and $0.20$ but not for the larger frustrations: for $\Jrat=0.60$ it stays at $\approx 0.2$, far below the exact value $1.58$.

At the abrupt change the entropy jumps to $\approx 2.6$--$3.2$, far above the exact value for every $\Jrat$, and the total spin jumps to $\ev{\mathbf{S}^2}\approx 2.7$--$3.4$, decreasing slowly to $\approx 2.2$--$2.4$ at $L=16$. Since the gates of the ansatz do not conserve $\mathbf{S}^2$, such a value means weight in $S>0$ sectors, which this ansatz can generate. The leakage out of the singlet sector is much stronger than for $N=10$, where $\ev{\mathbf{S}^2}$ peaked at $\approx 0.8$--$0.9$. The abrupt change of the energy error and of the fidelity is accompanied by an abrupt change of the entropy and of the total spin, so the four observables describe the same event in different ways.

\subsection{\texorpdfstring{$N=12$}{N=12}: the abrupt change is not a barren plateau}
\begin{figure}[t]
    \centering
    \includegraphics[width=0.95\columnwidth]{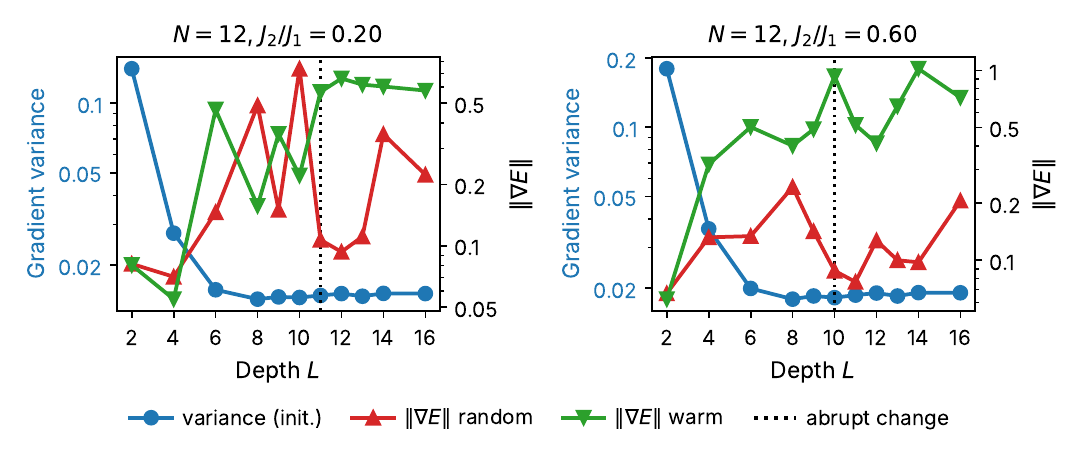}
    \caption{Gradient variance and $\|\nabla E\|$ versus depth ($N=12$). Gradient diagnostics for $N=12$ at $\Jrat=0.20$ (left) and $0.60$ (right). Blue: gradient component variance at random initialization. Red and green: $\|\nabla E\|$ at the optimized random and warm start runs. Dotted line: depth of the abrupt change.}
    \label{fig:N12grad}
\end{figure}

As for $N=10$, the abrupt change is not a barren plateau. Figure~\ref{fig:N12grad} shows the gradient-component variance at random initialization and $\|\nabla E\|$ at the optimized parameters for $\Jrat=0.20$ and $0.60$. After a drop at small depth, the variance stays constant at $\approx(1.4$--$2.0)\times10^{-2}$ for $L\ge 6$, with no decay with depth and no feature at the abrupt change ($L=11$ and $10$). At the optimized parameters, $\|\nabla E\|$ stays finite ($0.05$--$1.0$) both before and after the collapse. No run reaches $\|\nabla E\|<10^{-3}$ within $1500$ iterations, so the collapsed states are not shown to be converged minima.

\subsection{\texorpdfstring{$N=12$}{N=12}: the chained warm start}
The chained warm start removes the abrupt change for all six frustration values. In Figs.~\ref{fig:N12EF} and~\ref{fig:N12SS} the warm-start branch shows a decrease of the energy error and an improvement of the fidelity with depth, monotonic up to small frustration values $\Jrat=0.01$, $0.2$ and $0.4$, and its total spin stays close to zero, i.e.\ in the sector of the exact ground state.

For $\Jrat=0.01$ and $0.20$ the warm start reaches a fidelity of $\approx 0.99$ from $L=8$ on, with an energy error of $\approx 0.02$ at $L=16$ and an entropy that follows the exact value $\approx 1.2$. For $\Jrat=0.40$ the behavior is different: the warm start first stays on a plateau up to $L=11$ with fidelity $0.5$--$0.7$, energy error $\approx 0.15$--$0.3$, $\ev{\mathbf{S}^2}\approx 0$ and an entropy of only $\approx 0.1$--$0.3$, far below the exact value $1.23$. It escapes at $L=12$--$13$, where the energy error drops from $\approx 0.13$ to $\approx 0.06$ and $0.025$, the fidelity rises from $0.79$ to $0.99$ and the entropy rises from $\approx 0.5$ to the exact value. This escape occurs at $L=13$ for $N=12$, whereas for $N=10$ the fidelity was already $\approx 0.99$ at $L=8$. For $\Jrat=1.00$ the warm start is not better than the random branch at shallow depth: at $L=8$--$9$ the random branch has fidelity $\approx 0.91$ and the warm start only $\approx 0.45$--$0.47$, because the warm start inherits a poor solution from small $L$ (fidelity $\le 0.43$ up to $L=6$). In this case the optimizer reduces the energy error by reaching another minimum with a similar energy level.

\subsection{\texorpdfstring{$N=12$}{N=12}: failure window}
For $N=12$ the failure of the warm start is not restricted to a single frustration value as for $N=10$. It covers $\Jrat=0.60$, $0.80$ and $1.00$ (Fig.~\ref{fig:N12EF}); its lower edge lies between $0.40$ and $0.60$, close to the Majumdar--Ghosh point~\cite{Majumdar1969}, which is not among the values studied, and its upper end coincides with the largest value studied, so we cannot say where it ends. For $N=10$ the warm start reached a fidelity of $\approx 0.99$ at $\Jrat=0.80$ and $1.00$, so the window is wider for $N=12$.

Within the window the warm start does not reach the exact ground state. Its fidelity saturates at $\approx 0.50$ for $\Jrat=0.60$ (flat from $L=6$ on) and at $\approx 0.57$ for $\Jrat=0.80$ (flat from $L\approx 9$ on), and it is $\approx 0.57$ at $L=16$ for $\Jrat=1.00$, where it is still rising slowly. The energy error at $L=16$ is $\approx 0.05$, $0.20$ and $0.46$ for the three values. The entropy stays below the exact value ($\approx 0.25$ against $1.58$, $\approx 1.65$ against $1.78$, and $\approx 1.70$ against $1.95$), while the total spin stays in the $S=0$ sector ($\ev{\mathbf{S}^2}\lesssim 0.05$). The random branch lies on the same plateau up to $L=9$ and then collapses, and for $L\ge 10$ it never exceeds the warm start, in contrast to $N=10$, where it overtook the warm start at $L=16$ for $\Jrat=0.60$.

The case $\Jrat=0.60$ is the clearest example: the state has a small energy error ($\approx 0.05$) and $\ev{\mathbf{S}^2}\approx 0$, yet its fidelity is only $\approx 0.5$ and its entropy is far from the exact one. As for $N=10$, this shows that all conditions must be satisfied simultaneously to obtain the exact ground state, and that a state in the $S=0$ sector with a small energy error can match several properties of the ground state without being the ground state.

Exact diagonalization gives a first excitation gap of $\Delta\approx 0.08$ at $\Jrat=0.60$, against $\Delta\approx 0.41$, $0.37$ and $0.68$ at $\Jrat=0.40$, $0.80$ and $1.00$. The plateau at $\Jrat=0.60$ may therefore reflect the near-degeneracy of low-lying singlets. Its low entropy and $\ev{\mathbf{S}^2}\approx 0$ also resemble a dimerized (nearest-neighbor singlet-product) state, which for the half chain used here has $S_{\mathrm{vN}}=0$, an energy error of $\approx 0.32$ and an overlap of $\approx 0.41$ with the ground state; the observed plateau has a lower energy error and a somewhat higher fidelity, so it can at most be a dressed version of such a state. This should be tested with the exact eigenstate weights of the plateau state. For $\Jrat=0.80$ and $1.00$ the gap is not small, so near-degeneracy cannot explain the plateau there; it may reflect slow optimization (the fidelity at $\Jrat=1.00$ is still rising at $L=16$) or a different trap of the landscape. These data are consistent with, but do not prove, these explanations.

\subsection{Comparison of \texorpdfstring{$N=10$}{N=10} and \texorpdfstring{$N=12$}{N=12}}
Three differences stand out (Table~\ref{tab:compare}). (i) The abrupt change of the random branch shifts from $L\approx 9$ to $L\approx 10$--$12$ and becomes larger: the energy error reaches $\approx 10$ instead of $\approx 2.5$, $\ev{\mathbf{S}^2}$ reaches $\approx 3$ instead of $\approx 0.9$, and no recovery to high fidelity is seen up to $L=16$. (ii) The chained warm start still removes the abrupt change, but needs more layers to reach the ground state ($L=13$ instead of $L=8$ at $\Jrat=0.40$). (iii) The failure of the warm start grows from the single value $\Jrat=0.60$ to the range $0.60$--$1.00$.

\begin{table}[t]
\centering
\caption{Random-initialization collapse and warm-start failure for $N=10$ and $N=12$ (values read from the plots).}
\label{tab:compare}
\footnotesize
\setlength{\tabcolsep}{4pt}
\begin{tabular}{lcc}
\hline
 & $N=10$ & $N=12$ \\
\hline
Depth $L$ of the abrupt change & $\approx 9$ & $10$--$12$ \\
Parameters $P$ at the change & $\approx 180$ & $240$--$288$ \\
Peak energy error & $\approx 2.5$ & $\approx 7$--$10$ \\
Peak $\ev{\mathbf{S}^2}$ & $\approx 0.8$--$0.9$ & $\approx 2.7$--$3.4$ \\
Random fidelity at $L=16$ & $0.91$--$0.96$ & $0.14$--$0.52$ \\
Warm-start failure window & $\Jrat\approx 0.6$ & $0.6$--$1.0$ \\
Warm-start fidelity plateau & $\approx 0.58$ & $\approx 0.50$--$0.57$ \\
\hline
\end{tabular}
\end{table}

\subsection{Discussion}
Three points follow from the results. First, a symmetry-preserving ansatz can show a non-monotonic dependence on depth: the error improves, degrades abruptly at an intermediate depth, and recovers at larger depth. Because the family is nested, this is an optimization effect, and because the gradient stays finite it is not a barren plateau~\cite{McClean2018}. Second, the chained warm start follows a better branch at the same depth for most frustration values, which shows that the good state is accessible~\cite{Mele2022,Puig2026}. Third, the warm start is not universal: near $\Jrat=0.6$ for $N=10$, and over the window $\Jrat=0.6$--$1.0$ for $N=12$, it stays at a plateau although the energy and the total spin look acceptable. The comparison of four observables shows that no single one of them is sufficient.

The persistence of the failure under warm starting points to the structure of the low-energy spectrum and of the symmetry sectors accessible to the ansatz, not to the energy scale. Warm starting keeps the optimizer in the region of parameter space it entered at shallow depth. If that region leads to a state in a quasi-degenerate manifold or to a low-lying eigenstate of the wrong momentum or spin sector, more layers lower the energy without raising the fidelity. This is consistent with warm-start theory in which guarantees depend on a non-vanishing spectral gap~\cite{Puig2026}, and with the known tendency of circuits that break total-spin symmetry to drift between spin sectors, which symmetry-adapted variants restore~\cite{Seki2020}. The growth of both the collapse and the failure window from $N=10$ to $N=12$ suggests that these difficulties become more severe with system size, but two sizes do not establish a scaling law. These data are consistent with this picture but do not prove it.

\FloatBarrier
\section{Limitations and Outlook}\label{sec:limits}
\begin{itemize}
  \item The study covers $N=10$ and $12$ and the frustration values listed above; nothing is claimed outside them, and no scaling with $N$ is claimed from two sizes.
  \item All results use the N\'eel reference state and one ansatz family. Other ans\"atze, for example an SU(2)-symmetric exchange gate~\cite{Wiersema2020} or the present ansatz with a total-spin penalty, would show which conclusions are specific to SingleExcitationMinus.
  \item The simulations are noiseless and use exact statevectors.
  \item Near the Majumdar--Ghosh point the exact eigenstate weights and the spectral gap should be computed to test the near-degeneracy explanation. A stronger optimizer such as quantum natural gradient~\cite{Stokes2020} would test whether the plateau is an optimizer effect.
  \item Larger systems and two-dimensional lattices are next steps to study.
\end{itemize}

\section{Conclusion}\label{sec:concl}
We have shown for the periodic $J_1$--$J_2$ Heisenberg ring that a symmetry-preserving VQE ansatz with random initialization shows an abrupt change of the fidelity, the energy error, the entropy and the total spin at intermediate depth, followed by a gradual recovery for $N=10$, and that this is not a barren plateau. For $N=12$ the abrupt change occurs at larger depth and is much stronger, and no full recovery is seen up to $L=16$. A chained warm start removes the abrupt change for most frustration values, but it fails near $\Jrat=0.6$ for $N=10$ and over the window $\Jrat=0.6$--$1.0$ for $N=12$, where the fidelity saturates at a plateau of about $0.5$--$0.58$ although the state stays in the $S=0$ sector. Several conditions on the energy, the fidelity, the entropy and the total spin must therefore be satisfied simultaneously to identify the ground state. The results are limited to the system sizes and frustration values studied.

\section*{Acknowledgment}
S.~Sahoo acknowledges his family members for constant support. Part of the computation was performed at the Physics Department, Jadavpur University Computer Facility (Sanction No. SR/FST/PS-1/2022/219(C)).

\section*{Data and Code Availability}
The VQE code, the exact-diagonalization references, the data and the figure scripts used in this work are openly available at \textcolor{blue}{\url{https://github.com/soumyadip-sahoo/Heisenberg_J1_J2_VQE}}.

\hbadness=10000 

\end{document}